\documentclass[useAMS,usenatbib]{mn2e}

\usepackage[bookmarks,bookmarksopen,colorlinks,linkcolor={blue},citecolor={red},urlcolor={green}]{hyperref}
\usepackage{epsfig}
\usepackage[fleqn]{amsmath}

\def\be{\begin{equation}}
\def\ee{\end{equation}}

\def\gtsima{$\; \buildrel > \over \sim \;$}
\def\ltsima{$\; \buildrel < \over \sim \;$}
\def\prosima{$\; \buildrel \propto \over \sim \;$}
\def\gsim{\lower.5ex\hbox{\gtsima}}
\def\lsim{\lower.5ex\hbox{\ltsima}}
\def\simgt{\lower.5ex\hbox{\gtsima}}
\def\simlt{\lower.5ex\hbox{\ltsima}}
\def\simpr{\lower.5ex\hbox{\prosima}}
\def\la{\lsim}
\def\ga{\gsim}

\def\medea{{\tt MEDEA}}
\def\cosmomc{{\tt CosmoMC}}
\def\recfast{{\tt RECFAST}}
\def\camb{{\tt CAMB}}

\title[CMB constraints on light DM]{Cosmic microwave background constraints on light dark matter candidates}         
\author[C. Evoli, S. Pandolfi and A. Ferrara]
{C.~Evoli$^{1}$\thanks{E-mail: carmelo.evoli@desy.de}, S.~Pandolfi$^{2}$ and A.~Ferrara$^{3}$\\
$^1$ II. Institut f\"ur Theoretische Physik, Universit\"at Hamburg, Luruper Chaussee 149, D-22761 Hamburg, Germany\\
$^2$ Dark Cosmology Centre, Niels Bohr Institute, University of Copenhagen, Juliane Maries Vej 30, DK-2100 Copenhagen, Denmark\\
$^3$ Scuola Normale Superiore, Piazza dei Cavalieri 7, 56126 Pisa, Italy\\
}

\begin{document}
\maketitle
\label{firstpage}

\begin{abstract}
Unveiling the nature of cosmic dark matter (DM) is an urgent issue in cosmology. Here we make use of a strategy based on the search for the imprints left on the cosmic microwave background temperature and polarization spectra by the energy deposition due to  annihilations of the most promising DM candidate, a stable weakly interacting massive particle (WIMP) of mass $m_\chi =1-20$ GeV. A major improvement with respect to previous similar studies is a detailed treatment of the annihilation cascade and its energy deposition in the cosmic gas.
This is vital as this quantity is degenerate with the annihilation cross-section $\langle \sigma v \rangle$. 
The strongest constraints are obtained from  Monte Carlo Markov chain analysis of the combined WMAP7 and SPT data sets up to $\ell_{\rm max} = 3100$. If annihilation occurs via the $e^+-e^-$ channel, a light WIMP can be excluded at the 2 $\sigma$ confidence level as a viable DM candidate in the above mass range. However, if annihilation occurs via $\mu^+-\mu^-$ or $\tau^+-\tau^-$ channel instead
%, differently from previous simplified studies 
we find that WIMPs with $m_\chi > 5$ GeV might represent a viable cosmological DM candidate. 

We compare the results obtained in this work with those obtained adopting an analytical simplified model for the energy deposition process widely used in the literature, and we found that realistic energy deposition descriptions can influence the resulting constraints up to 60\%.

\end{abstract}


%%%%% INTRO %%%%%

\section{Introduction}

According to the widely accepted $\Lambda$ Cold Dark Matter ($\Lambda$CDM) cosmology, the Universe is mostly made of {\it dark} components, i.e. dark energy (75\% of the mass-energy budget) and dark matter (DM; 20\%); these components largely dominate over baryons \citep{Komatsu:2011}. 
The situation is then rather unsatisfactory as the nature of the dark components is far from being established and it stands as one of the most crucial issues in cosmology.

The most promising DM interpretation is in terms of a thermal relic density of stable weakly interacting massive particles (WIMPs). 
An appealing feature of such a scenario is that the annihilation cross-sections predicted by the electroweak scale automatically provide the right DM density after freeze-out~\citep{Bertone:2010}.  This argument applies equally well to particles with $1-20$~GeV masses as to those with masses more traditionally associated with supersymmetric neutralinos ($m_{\chi} \sim 40-1000$~GeV).

In the recent years, pieces of evidence have been accumulating in favour of DM in the form of $\sim\!10$~GeV WIMPs. 
In fact, a relatively light DM particle with an annihilation cross-section consistent with that predicted for a simple thermal relic ($\langle \sigma v \rangle_T  \sim 10^{-26}$~cm$^{3}$ s$^{-1}$) and a distribution in the halo of the Milky Way consistent with that predicted from simulations could accommodate the indirect detection of gamma-rays from the Galactic Centre, the synchrotron emission from the Milky Way radio filaments and the diffuse synchrotron emission from the inner galaxy (the so-called ``WMAP Haze''\footnote{Notice that the Haze has been confirmed by Planck and appears to be spatially coincident with the Fermi bubbles which suggests a non-DM explanation for at least a substantial portion of the emission\citep{Su:2010,Hooper:2013}.} \citep{Finkbeiner:2004,Hooper:2007,Hooper:2011b,Hooper:2011,Dobler:2010}. 

At the same time it would be compatible with claims of low-energy signals from DM direct detection experiments as DAMA/LIBRA, CoGeNT, and CRESST-II. 
In particular, the striking detection of annual modulation observed by DAMA/LIBRA (now supported by CoGeNT) appears inconsistent with all known standard backgrounds. 
Note, however, that (a) other experiments, such as CDMS and XENON100, have not confirmed the result of the direct detections, and (b) indirect detection features might have alternative astrophysical explanations 
\citep{Bernabei:2008,Biermann:2010,Akerib:2010,Ahmed:2010,Bernabei:2010,Crocker:2011,Aalseth:2011,Aalseth:2011b,Aprile:2011,Xenon100:2012,Guo:2012}.

A phenomenological model of light DM particle able to accommodate the collection of indirect and direct observations should require that DM annihilates primarily into leptons with a cross-section close to $\langle \sigma v \rangle_T$. 
Moreover, approximately 20\% of annihilations must also proceed to hadronic final states in order to yield a spin-independent, elastic scattering cross-section ($\approx 10^{-41}$~cm$^{2}$) with nucleons compatible with the direct detection~\citep[see][for a detailed review]{Hooper:2012}.

The light DM hypothesis implies a larger cosmic number density of such particles ($n_{\rm DM} \propto \Omega_{\rm DM} h^{2}/m_{\rm DM}$); in addition, the annihilation rate ($\propto n_{\rm DM}^{2} \propto (1+z)^{6}$) increases dramatically at early cosmic times. 
These two facts imply that the annihilation energy deposition might profoundly affect the thermal and ionization history of the intergalactic medium (IGM)\footnote{Strictly speaking, the term ``intergalactic medium'' is ill-defined before the epoch of galaxy formation. Nevertheless, following common practice, we will use it anyway.} prior to reionization. 
In turn, this modified evolution with respect to the  standard recombination scenario can in principle leave detectable signatures in the cosmic microwave background (CMB) anisotropy power spectrum\footnote{Electrons, positrons and photons deposit their energy very efficiently in the IGM; hence leptonic channels are expected to leave a stronger CMB signature}. 
Determining the amplitude of this effect is the chief goal of the present study. 
 
The effects of the DM annihilation around the redshift of the last scattering surface (LSS) have been discussed in~\citet{Padmanabhan:2005} and are only briefly summarized here. 
The extra free-electrons resulting from the DM energy cascade scatter CMB photons, thus thickening the LSS and in principle shifting the position of the peaks in the temperature-temperature (TT) power spectrum.
In practice, reasonable electron density excesses yield corrections to the positions of the peaks that can be safely ignored here. 
More importantly, oscillations on scales smaller than the LSS width are damped in the TT and EE spectra in a manner inversely proportional to their wavelength.  
Such DM annihilation effects on the TT spectrum are degenerate with variations of the slope ($n_s$) and amplitude ($A_s$) of the primordial power spectrum, and, to a lesser extent, with the baryon ($\Omega_bh^2$) and DM ($\Omega_{\rm DM}h^2$) density parameter. 
Polarization spectra are generated via Thomson scattering of the local quadrupole in the temperature distribution. 
As the broadening of the LSS increases the intensity of the quadrupole moment, the EE spectrum is enhanced on large scales. 
Furthermore, it can be shown \citep[i.e.][]{Padmanabhan:2005} that the quadrupole is dominated by the free-streaming from the dipole perturbation that is $\pi/2$ out of phase of the monopole. 
A thicker LSS boosts the fractional contribution from the monopole, thus slightly shifting the peaks of the EE and TE spectra. 

A key aspect of these calculations is that only a fraction of the released energy is finally deposited into the IGM in the form of heating and H/He ionizations. 
However, earlier studies~\citep{Padmanabhan:2005,Mapelli:2006,Galli:2009} have used a simplified description of such processes, based on the hypothesis that a redshift-independent fraction of the DM rest-mass energy is absorbed by the IGM.
More recently,~\citet{Slatyer:2009,Galli:2011,Hutsi:2011} have reassessed the energy deposition problem including various energy-loss mechanisms in a more realistic way. 
This approach, based on semi-analytical solutions lacks an implementation of low-energy atomic processes that determine the actual absorption channel (e.g. heating, ionization, excitations) and because of this they have to rely on the results of~\cite{Chen:2004}.
To fill this gap here we build upon our previous work~\citep{Valdes:2010} in which we developed the Monte Carlo Energy Deposition Analysis ({\tt MEDEA}) code which includes bremsstrahlung and inverse Compton processes, along with H/He collisional ionizations and excitations, and electron-electron collisions. 
{\tt MEDEA}  enables us to compute the energy partition into heating, excitations and ionizations as a function of the primary initial energy, the gas ionization fraction and the redshift. 
{\tt MEDEA} has been recently improved \citep{Evoli:2012} to include the energy cascade from particles generated by primary leptons/photons using the most up-to-date cross-sections and extending the validity of the model to unprecedented high ($\sim$~TeV) energies~\citep[see][]{Shull:1979,Shull:1985,Furlanetto:2010}. 
In addition, arbitrary initial energy distribution of electrons, positrons and photons can be assigned. 
These improvements make {\tt MEDEA} suitable for studying the IGM energy deposition for some of the most popular DM candidates~\citep{Evoli:2012}. 
With this greatly improved physical description we aim at computing the signatures left in the CMB spectrum by annihilating light DM.

%%%%% METHOD %%%%%

\section{Method}

In this Section we compute the energy input of DM annihilations in the IGM. This approach is similar in spirit to a number of recent  works~\citep{Padmanabhan:2005,Galli:2009,Hutsi:2009,Slatyer:2009,Galli:2011,Hutsi:2011,Natarajan:2012}; however, we improve upon them by a more accurate description of the energy deposition channels.

\subsection{Modified ionization history}\label{Sec:medea}
\begin{figure}
\centering 
\includegraphics[width=0.49\textwidth]{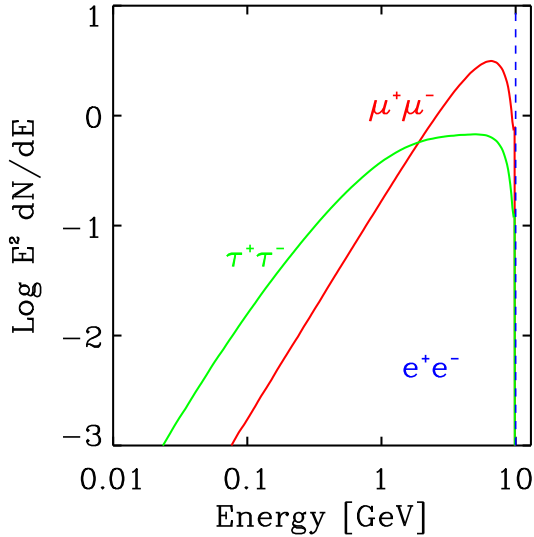}
\caption{Energy spectrum of electrons or positrons from the annihilation of a 10 GeV mass WIMP into $e^+e^-$, $\mu^+\mu^-$ and $\tau^+\tau^-$ channels.\label{Fig:spectra}}
\end{figure} 
For the reasons given in the Introduction, we concentrate on light DM candidates that annihilate mainly in leptonic channels. 
In Fig.~\ref{Fig:spectra} we show the annihilation spectra of a 10~GeV DM particle for the different annihilation channels, computed using the public code {\tt DarkSUSY}. 
The muonic and tauonic channels produce a leptonic pair
whose prompt annihilation gives rise to an energy spectrum of primary electrons or positrons with kinetic energy from 10~GeV down to few tens of MeV; 
annihilation in the electron channel produces an electron/positron pair in which both the two primary leptons have a kinetic energy which is the mass of the annihilation particle. 

The total energy density input from DM annihilations is:
\begin{multline}\label{Eq:release}
\frac{dE_{\rm DM}}{dt} (z) = \rho_{c}^{2} c^{2} \Omega_{\rm DM}^{2} (1+z)^{6} \frac{\langle \sigma v \rangle}{m_{\rm DM}} \\
\approx 4.03 \times 10^{-38} \!\left( \! \frac{\Omega_{\rm DM}h^{2}}{0.11} \! \right) (1+z)^{6} B  \left( \!\frac{m_{\rm DM}c^{2}}{\rm GeV} \!\right)^{\!-1} \, {\rm GeV \, cm^{-3} \, s^{-1}} 
\end{multline}
where $\rho_{c}=3 H_{0}^{2}/ 8 \pi G$ is the critical density of the universe today, $\Omega_{\rm DM}$ is the DM density contribution to the critical density, $m_{\rm DM}$ is the mass of the DM 
particle and $\langle \sigma v \rangle$ is the thermally averaged product of the cross-section and relative velocity of the annihilating DM particles. Moreover we have defined 
$B \equiv \langle \sigma v \rangle / 3 \times 10^{-26}$~cm$^{3}$~s$^{-1}$. Note that equation~\ref{Eq:release} is valid only for DM Majorana particles.

In the light of the earlier works of~\cite{Cirelli:2009} and \cite{Hutsi:2011} we neglect the role of structure formation in the calculation of the energy deposition. 
In fact, haloes with density higher than the background could in principle boost the average annihilation rate; however, their formation starts at a relative low redshift ($z\lsim 100$) when 
the ionization rate due to DM annihilation is already negligible. 

By introducing the mean number density of hydrogen nuclei $n_{\rm H} \approx 1.9 \times 10^{-7} (1+z)^{3}$~cm$^{-3}$ and the parameter  
\begin{equation}
\epsilon_{0} \equiv 2.12 \times 10^{-31} \left( \! \frac{\Omega_{\rm DM}h^{2}}{0.11} \! \right) B  \left( \!\frac{m_{\rm DM}c^{2}}{\rm GeV} \!\right)^{\!-1} \, ,
\end{equation} 
equation~\ref{Eq:release} becomes:
\begin{equation}
\frac{dE_{\rm DM}}{dt} (z) = \epsilon_{0} n_{\rm H} (1+z)^{3} \, {\rm GeV \, s}^{-1}
\end{equation}

It has been pointed out that $\langle \sigma v \rangle$ could be somewhat boosted by the Sommerfeld effect \cite[e.g.][]{Galli:2009,Slatyer:2009}. 
Although it is easy to implement this process in this scheme we have not considered it here as it depends strongly on the DM model chosen~\citep{vandenAarssen:2012}. Moreover, over the parameter space considered by most studies, this effect can also be approximated as a constant boost to the annihilation rate over the redshift range of interest and then applied to our final results.

To derive the DM-modified cosmic ionization/thermal history, we need to include the above heating (and corresponding ionization) rate into the relevant detailed balance equations. To this aim, we have modified the publicly available code\footnote{Recently other similar codes (e.g. CosmoREC, HyREC) have improved the precision of the results implementing a more detailed description of the atomic structure~\citep{Hutsi:2011,Giesen:2012}. However, given the current precision of CMB experiments, these corrections do not affect our conclusions.} \recfast~\citep{Seager:1999}, part of the \camb~\citep{Lewis:1999bs} code, by adding the following terms:
\begin{align}\label{Eq:extra1}
-\frac{dx_{\rm H}}{dz} & = \frac{1}{H(z)(1+z)} f_{\rm ion,H}(z) \frac{dE_{\rm DM}/dt}{n_{H}(z)E_{\rm ion,H}} \\
-\frac{dx_{\rm He}}{dz} & = \frac{1}{H(z)(1+z)} f_{\rm ion,He}(z) \frac{dE_{\rm DM}/dt}{n_{H}(z)E_{\rm ion,He}} \\
-\frac{d T_{M}}{dz} & = \frac{1}{H(z)(1+z)}\frac{2}{3k_{B}} \frac{f_{h}(z)}{1+f_{\rm He}+x_{e}(z)}\frac{dE_{\rm DM}/dt}{n_{H}(z)}\label{Eq:extra3}
\end{align}
where $f_{\rm ion,H}$, $f_{\rm ion,He}$ ($f_{\rm h}$) are energy deposition fractions into H or He ionizations (heating) including those induced by Ly$\alpha$ photons on atoms in the excited states.

A key point to take from equations~\ref{Eq:extra1}$-$\ref{Eq:extra3} is that energy deposition fractions are fully degenerate with the parameter we aim to constrain, i.e. $\langle \sigma v \rangle$. To partly alleviate this difficulty, a possible strategy, first proposed by~\citet{Ripamonti:2007}, is to determine the lowest possible bound by assuming
$f_{h}=f_{\rm ion}=1$. More often, constraints have been derived by using the \cite{Chen:2004} prescription for $f_i$.  Based on the results of~\cite{Shull:1985}, these authors pointed out that when the gas is mostly neutral, energy is evenly distributed among ionizations, excitation and heating; for a fully ionized medium, almost all of the energy goes instead into gas heating. A linear interpolation is used for intermediate ionization values:
\begin{eqnarray}\label{Eq:Chen}
f_i &=&  \frac{1}{3}(1- x_e) \\
f_h &=& \frac{1}{3} (1 + 2x_e) \, .
\end{eqnarray}
This approximation is too crude to be used for high-precision predictions as clearly shown by the comparison with fully fledged Monte Carlo simulations~\citep{Valdes:2007,Furlanetto:2010}. Moreover, for primary energies $\simgt 1$ MeV, inverse Compton energy losses on the CMB become important and introduce a significant redshift dependence of the fractions.
These processes have been carefully modelled in~\cite{Evoli:2012} and here we use their results for $f_i$. Note that the latter assume that photon energy deposition occurs locally, which is not true in general~\citep[see][]{Slatyer:2009}. In Appendix~\ref{Appendix:A}, we show that this approximation is very accurate in the energy range of interest here.
\begin{figure}
\centering 
\includegraphics[width=0.49\textwidth]{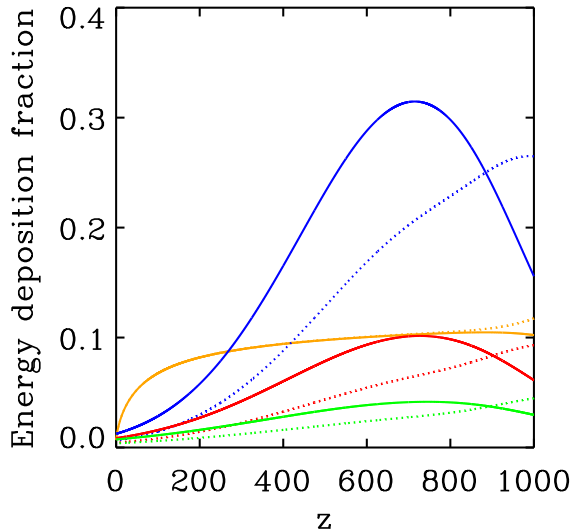}
\caption{Fractional energy depositions into H and He ionization (solid lines) and heating (dotted lines) from DM annihilation of a 10 GeV DM candidate as a function of redshift. 
Colour code as in Fig.~\ref{Fig:spectra}. The orange lines show the energy deposition fractions obtained from the simplified deposition model described in Section~\ref{comparison} for annihilation into muons. In this plot, we assume the standard IGM ionization evolution based on a WMAP7 cosmology.\label{Fig:deposition}}
\end{figure} 
The energy depositions calculated with MEDEA are shown in Fig.~\ref{Fig:deposition} for different annihilation channels of a 10 GeV DM particle mass. Such curves show a dependence on the annihilation channel since different initial spectral distributions involve different energy loss mechanisms. For computational speed-up purposes, we have derived handy fitting formulae, given in Appendix B, to the {\tt MEDEA} numerical results. 

\subsection{MCMC analysis}\label{Sec:cosmomc}

To obtain a constraint on the annihilation cross-section of light DM candidate, we have performed a Monte Carlo Markov chain (MCMC) analysis using the publicly available \cosmomc~package \citep{Lewis:2002}. 
We consider here a flat $\Lambda$CDM model with the canonical six parameters plus an additional seventh one, $\langle \sigma v \rangle$. 
Therefore, the theoretical model we adopt is described by the following set of parameters:
\begin{equation}
\{\omega_b,\omega_{\rm DM}, \theta_s, \tau, n_s, \log[10^{10}A_{s}], \langle \sigma v \rangle\},
\label{parameter}
\end{equation}
where $\omega_b\equiv\Omega_bh^{2}$ and $\omega_{\rm DM}\equiv\Omega_{\rm DM}h^{2}$ are the baryons and CDM density parameters, $\theta_{s}$ is the ratio between the sound horizon and the angular diameter distance at decoupling, $\tau$ is the optical depth, $n_s$ is the scalar spectral index and $A_{s}$ is the amplitude of the primordial spectrum. 
The flat priors assumed for these parameters are shown in Tab.~\ref{tab:priors}.
\begin{table}
\begin{center}
\begin{tabular}{c|c}
\hline\hline
Parameter & Prior\\
\hline
$\Omega_{b}h^2$ & 0.005-0.1\\
$\Omega_{\rm DM}h^2$ & 0.01-0.99\\
$\theta_s$&0.5-10\\
$\tau$ & 0.01-0.8\\
$n_{s}$ & 0.5-1.5\\
$\ln{(10^{10} A_{s})}$ & 2.7-4\\
$\langle \sigma v \rangle$/cm$^{3}$~s$^{-1}$ & 0 - $10^{-24}$\\
\hline\hline
\end{tabular}
\caption{Adopted flat priors for the cosmological parameters.}
\label{tab:priors}
\end{center}
\end{table}
Our basic data set is the 7--yr WMAP temperature and polarization data~\citep{Komatsu:2011,Larson:2011}. We consider purely adiabatic initial conditions and we impose spatial flatness. We also fixed the primordial fractional abundance of helium to the standard observed nominal value of $Y_{\rm He}=0.24$. 
We refer to this basic data set as ``\emph{WMAP7}''.  For each case we run five chains; convergence diagnostic tests are performed using the Gelman and Rubin ``variance of chain mean/mean of chain variances" R-1 statistics. We consider the chains to be converged only if $R-1 < 0.03$. The 68 and 95 per cent confidence level (c.l.) one- and two-dimensional constraints are obtained after marginalization over the remaining ``nuisance'' parameters. We have tested that varying H$_{0}$ instead of $\theta_s$, as suggested in~\cite{Galli:2009}, our results are found to be affected by less than 5 per cent.

In addition to the \emph{WMAP7} data set we also consider the case ``\emph{CMB ALL+SPT}''. In this larger data set we include, in addition to the WMAP data, the CMB temperature and polarization data from QUaD~\citep{Brown:2009uy}, and the recent SPT~\citep{Keisler:2011aw} data.  The inclusion of the QUaD experiment (a) enlarges the multipole range considered for the temperature, allowing us to probe the small-scale region $500\le \ell \le 2500$, and (b) adds information on the $E$- and $B$-mode polarization. Moreover, the SPT experiment pushes the dynamic range of CMB observations to larger multipoles with the respect of WMAP7, measuring with a better accuracy the damping tail of the CMB angular power spectrum.
We consider data up to $\ell=3100$. For the SPT experiment, it is necessary to account for foreground contributions by adding three extra parameters representing the amplitude of the SZ, $A_{\rm SZ}$, clustering, $A_{C}$, and shot-noise, $A_{P}$, signal from point sources. We used for each foreground component the proper template provided by~\citet{Keisler:2011aw}. When deriving our constraints we marginalize over these three nuisance parameters.
To compute the likelihood of the data we have properly modified the \cosmomc~package in order to make use of the
routine supplied by the WMAP team for the WMAP7 data set, publicly available at the LAMBDA website\footnote{http://lambda.gsfc.nasa.gov/}, and of the likelihood code provided by the SPT team~\citep{Keisler:2011aw} for the SPT experiment.

As we already discussed in the introduction, the inclusion of small-scale CMB measurements can greatly help in breaking the degeneracy with the other cosmological parameters, and in particular with $n_s$, thus improving the constraints on the DM sector parameters. Moreover, the addition of the SPT data to the WMAP data improves the constraints on the ratio of the sound horizon to the angular diameter distance parameter $\theta_s$ by nearly a factor of 2~\citep{Keisler:2011aw}, thus narrowing the allowed range of the other parameters.

We adopt the standard parametrization for the reionization, considered as an instantaneous process occurring at some redshift $z_r$, with $z_{r}<32$. Such a choice leads to a one-to-one relation between $z_r$ and the adopted e.s. optical depth $\tau$. As a caveat, we note that \citet{Pandolfi:2012} showed that a more realistic reionization modelling might affect the cosmological parameters that are more degenerate with the DM annihilation cross-section, thus introducing an additional source of uncertainty\citep[see however also][]{Moradinezhad:2012}. 

%%%%%% RESULTS

\section{Results}
\begin{figure}
\centering 
\includegraphics[width=0.49\textwidth]{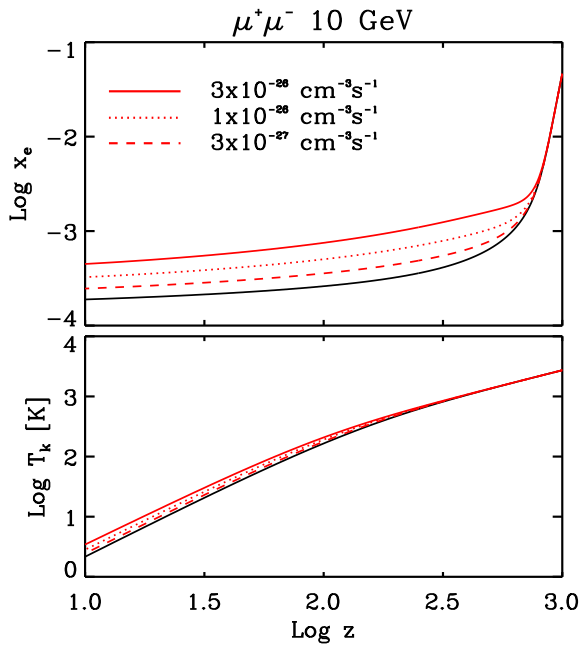}
\caption{DM-modified ionization (top) and thermal (bottom) histories for a 10 GeV WIMP annihilating into muons. Values of the annihilation cross-section correspond to different curves as shown by the legend. The black solid line represents the case without DM annihilations.\label{Fig:xe}}
\end{figure} 
\begin{figure}
\centering 
\includegraphics[width=0.49\textwidth]{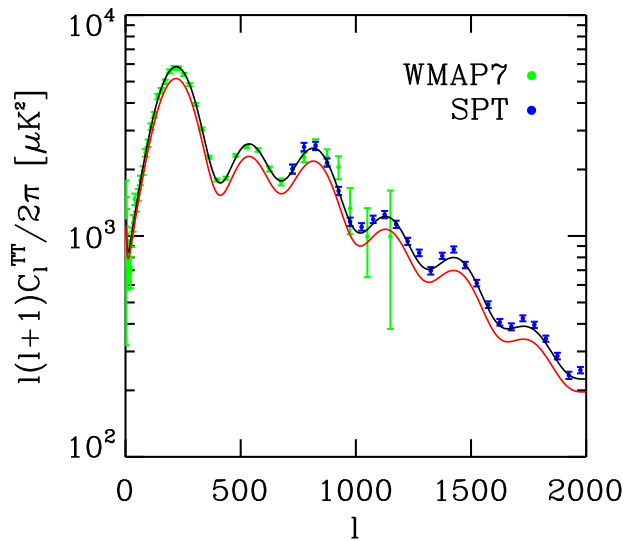}
\caption{Angular power spectrum of CMB temperature fluctuations: standard case without DM annihilations (black line), 
considering a 10 GeV WIMP annihilating into $\mu^+-\mu^-$ with $\langle \sigma v \rangle = 3\times 10^{-26}$~cm$^{3}$~s$^{-1}$ (red line). 
The points with errorbars show the 7-yr measurements of the WMAP satellite (black) and the SPT data (blue).\label{Fig:Cls}}
\end{figure} 
\begin{figure}
\centering 
\includegraphics[width=0.49\textwidth]{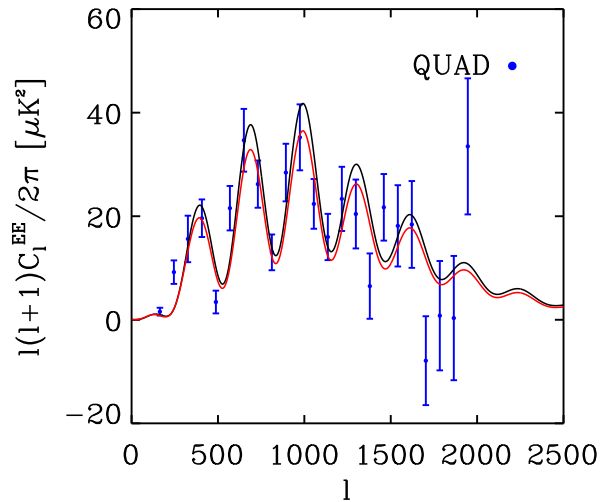}
\caption{As in Fig.~\ref{Fig:Cls} for the polarization fluctuations.The points with error bars show the measurements of the QUaD experiment.\label{Fig:Cls_ee}}
\end{figure} 
DM-modified ionization and thermal histories for a 10 GeV WIMP annihilating into muons on top of a $\Lambda$CDM model are shown in Fig.~\ref{Fig:xe}; the corresponding TT (EE) spectrum is shown in Fig.~\ref{Fig:Cls} (Fig.~\ref{Fig:Cls_ee}). Qualitatively similar conclusions can be drawn for the other channels.
The energy released in the form of electrons and positrons from the annihilation of DM particles delays and quenches the recombination processes, thus resulting in a freeze-out relic electron fraction a factor of a few larger, depending on the value of $\langle \sigma v \rangle$. 
For the same reason, the temperature drop with time is less pronounced. As a consequence of the higher ionization rate, the CMB normalization value is smaller. 

\begin{figure*}
\centering 
\includegraphics[width=0.49\textwidth]{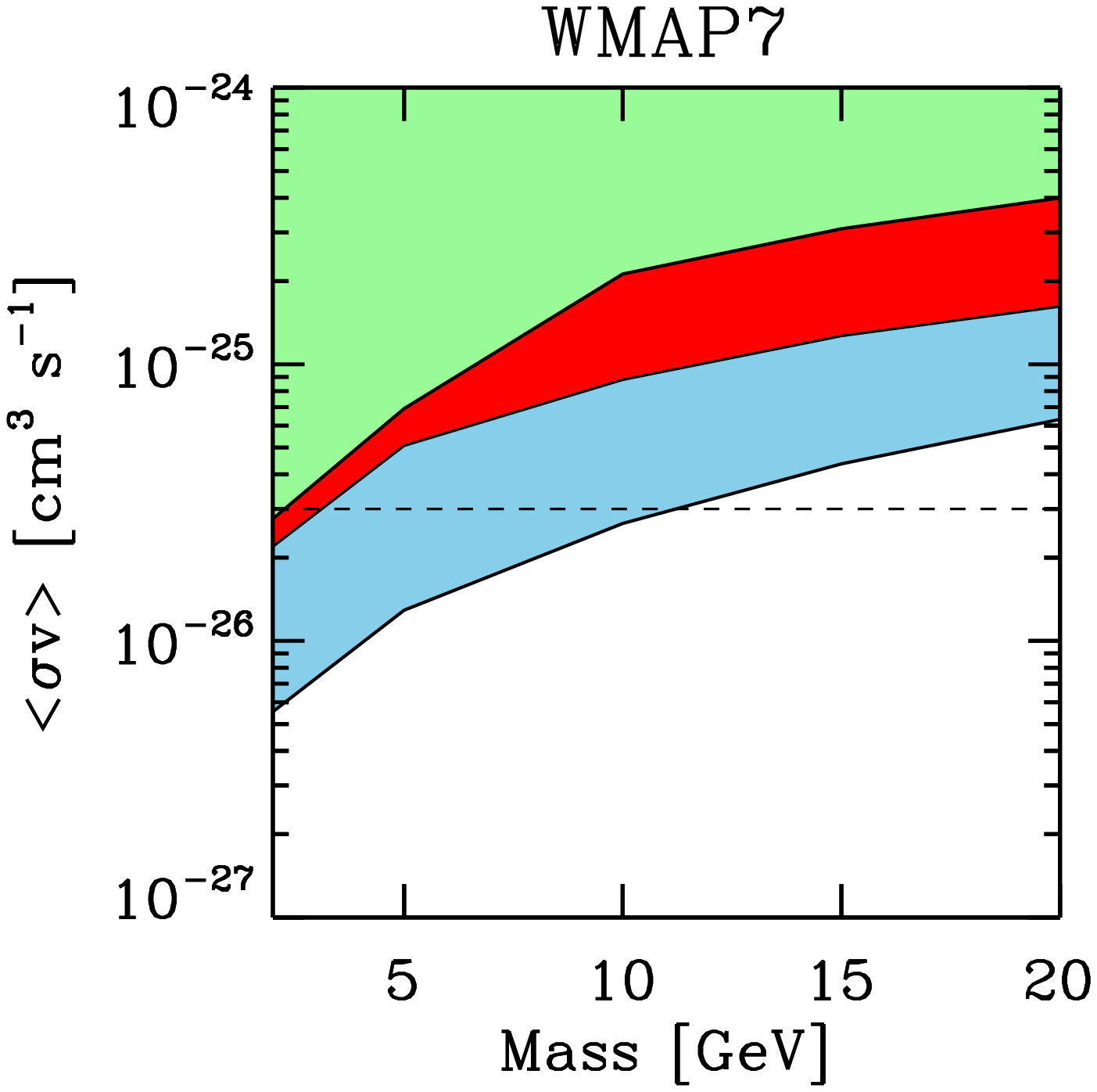}
\includegraphics[width=0.49\textwidth]{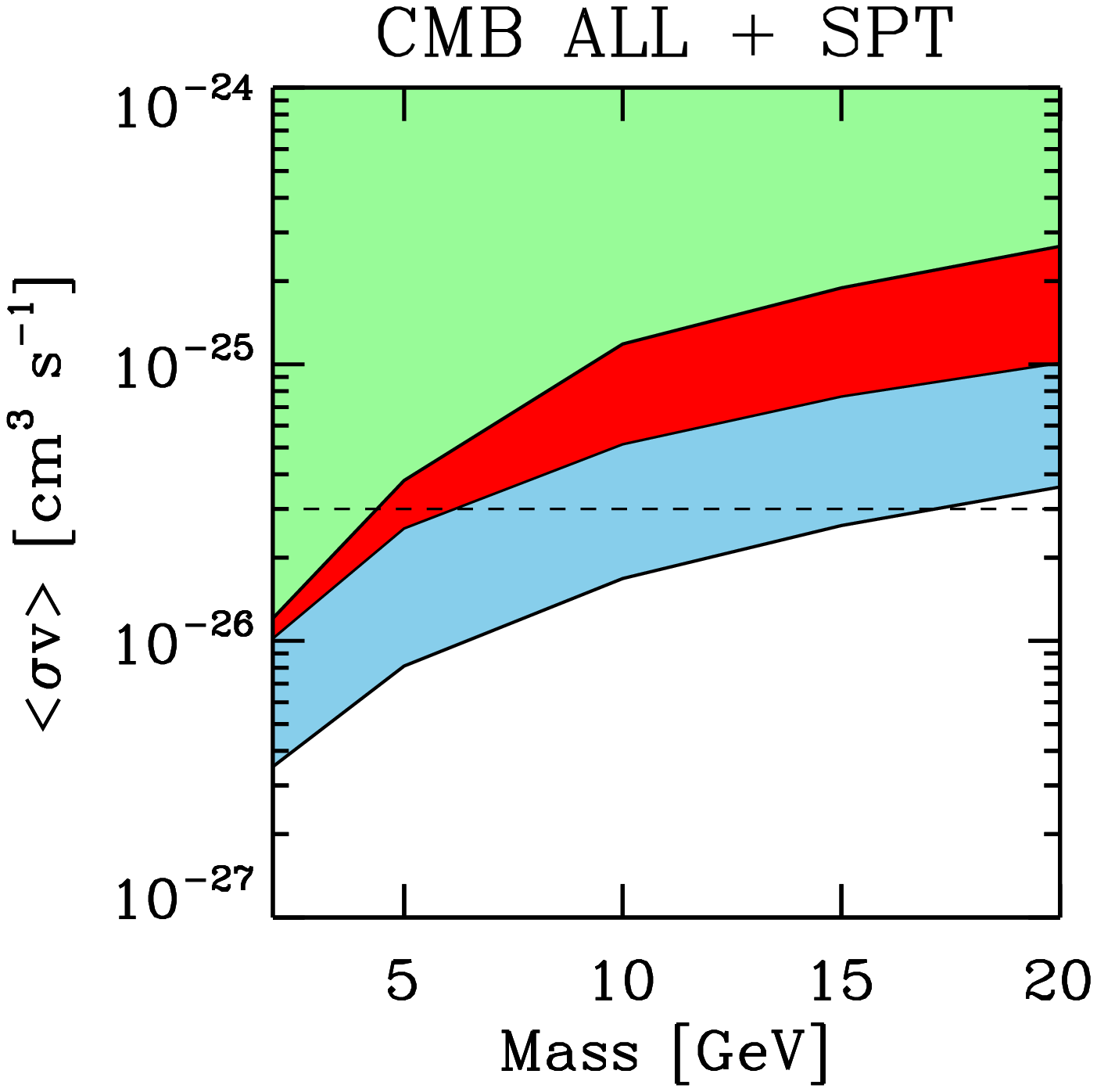}
\caption{Constraint plot on the maximum cross-section for different DM candidates based on \emph{WMAP7} and \emph{CMB ALL+SPT} data set 
(the region excluded for the tau annihilation channel is indicated in green, while the additional region excluded for the muon (electron) annihilation channel is indicated in red (blue)).\label{Fig:constraints}}
\end{figure*}

We follow the procedure described in Section~\ref{Sec:cosmomc} to get constraints on the cosmological parameters in equation (\ref{parameter}) and we compare them with those obtained by the WMAP team from their 7-yr data. We present in Fig.~\ref{Fig:constraints} the 2 $\sigma$ c.l. constraints on the DM annihilation cross-section $\langle \sigma v \rangle$ as a function of the DM mass. 
Differently to~\citep[e.g.][]{Galli:2009} our results cannot be given as a single number due to the mass dependence of the energy deposition fractions (see Section~\ref{Sec:medea}). A detailed comparison with their results will be given in Section~\ref{comparison}. 
The main conclusion is that only DM candidates lighter than $\le 10$~GeV annihilating via the $e^+-e^-$ channel can be excluded as a dominant component of the DM energy density.   
The constraints are stronger, as expected, if we include in the present analysis the recent SPT data set with $\ell_{\rm max} = 3100$\footnote{We have verified that our results are insensitive for the choice of $3000 \lsim \ell_{\rm max}\lsim 3500$.} and the polarization data. In this case the electron channel is excluded in the entire mass range (up to 20 GeV), where the other two channels can be excluded for masses $\lsim5$~GeV. 

We have verified that the stronger constraints come mainly from the SPT data inclusion, since the polarization data alone improve the constraints by $< 3$\%.  Currently polarization data alone are not of sufficient quality to robustly constrain DM parameters. Future experiments, specifically devoted to measure polarization at smaller scales like Planck~\citep{Tauber:2010}, PolarBear~\citep{Anthony:2012} and CMBPol~\citep{Zaldarriaga:2008} are expected to significantly  improve the situation.
\begin{figure*}
\centering 
\includegraphics[width=\textwidth]{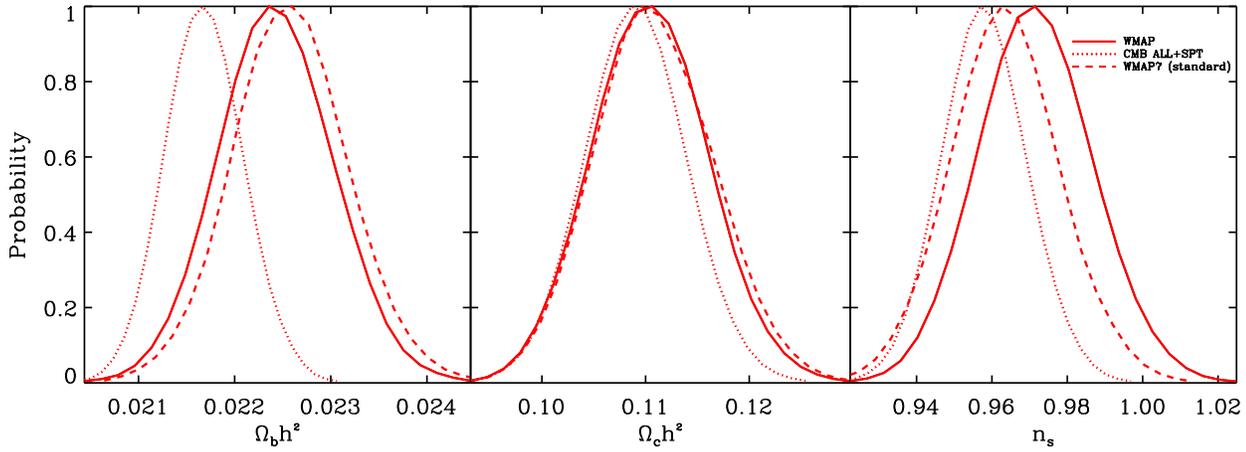}
\caption{One dimensional posterior probability distribution of $\Omega_{b}h^{2}$, $\Omega_{\rm DM}h^{2}$ and $n_{s}$ parameters in the case of WMAP7 data set and no DM annihilation (dashed), \emph{WMAP7} case (solid) and \emph{CMB ALL + SPT} case (dotted)\label{Fig:probability}}
\end{figure*} 
\begin{table*}
\caption{95\% c.l. upper limit constraints on the DM annihilation cross-section $\langle \sigma v \rangle$ [$10^{-26}$~cm$^{3}$/s] in different cases of mass, annihilation channel and data set considered.}\label{tab3}
\begin{center}
\small
\begin{tabular}{cccc|ccc}
\hline\hline
Mass     & \multicolumn{3}{c|}{WMAP7} &\multicolumn{3}{c}{CMB ALL+SPT}\\
\hline
             & $e^+e^-$&$\mu^+\mu^-$&$\tau^+\tau^-$& $e^+e^-$&$\mu^+\mu^-$&$\tau^+\tau^-$\\
\hline
2 GeV   & $\leq$0.554 & $\leq$2.19 & -       & $\leq$0.351& $\leq$1.02 &  - \\
5 GeV   & $\leq$1.29   & $\leq$5.07 & $\leq$6.91 &$\leq$0.811& $\leq$2.55 &$\leq$3.80\\
10 GeV & $\leq$2.66   & $\leq$8.79 & $\leq$21.2 &$\leq$1.68  & $\leq$5.14 & $\leq$11.8\\
20 GeV & $\leq$6.32   & $\leq$16.2 & $\leq$40.0 & $\leq$3.60  &$\leq$10.1 & $\leq$26.7\\
\hline\hline
\end{tabular}
\end{center}
\label{default}
\end{table*}
\begin{table*}
\begin{center}
\begin{tabular}{c|ccc}
\hline\hline
Parameter & WMAP7 (Standard) & WMAP7  & CMB ALL+SPT \\
\hline
$\Omega_{b}h^2$&  0.0226 $\pm$ 0.0006& 0.0224$\pm$0.0006& 0.0217$\pm$ 0.0004\\
$\Omega_{DM}h^2$ &0.1109$\pm$  0.0056&0.1105$\pm$  0.0054&0.1099 $\pm$ 0.0049 \\
$\theta_s$&1.0388$\pm$ 0.0027&1.0384$\pm $ 0.0027& 1.0397$\pm$0.0015\\
$\tau$ &0.0884$\pm$ 0.0152&0.0876 $\pm$ 0.0150&0.0831$\pm$  0.0138 \\
$n_{s}$ &0.9635$\pm$ 0.0142&0.9716 $\pm$ 0.01479& 0.9579$\pm $ 0.0113\\
$\ln{(10^{10} A_{s})}$ &3.1904$\pm$ 0.0457& 3.1868$\pm$0.0460 &  3.2122  $\pm$ 0.0446\\
\hline\hline
\end{tabular}
\caption{68\% c.l. constraints of the background cosmology parameters in the case of WMAP7 data set with no DM annihilation (WMAP7 (standard)), compared with the WMAP7 case with DM annihilation, and the CMB ALL+SPT case.}  
\label{tab4}
\end{center}
\end{table*}
The CMB constraints we find are weaker than the constraints obtained by the Fermi experiment using the signal in the diffuse isotropic gamma emission from the Galaxy~\citep{Abdo:2010c} and from a combined analysis of the Milky Way satellites~\citep{Ackermann:2011,Baushev:2012,Cholis:2012}. Comparing the 10 GeV case of annihilation channel in muons and that in taus, the inferred maximum cross-section from Fermi falls below the thermal value. However, in their analysis the rather uncertain distribution of DM in galaxies must be specified, while the present approach is free from any such hypothesis.

In Table \ref{tab4} we report the 68\% c.l. constraints on the cosmological parameters for the 10 GeV muon annihilation channel for the \emph{WMAP7} and \emph{CMB ALL+SPT} cases, and the WMAP7 alone data set, i.e. a minimal $\Lambda$CDM model without annihilating DM (``WMAP7 (Standard)''). The one-dimensional posterior probability for $\Omega_bh^2$, $\Omega_{\rm DM}h^2$ and $n_s$ for the three data set cases considered is also shown in Fig.~\ref{Fig:probability}. The strongest shift occurs for the baryon density $\Omega_bh^2$ which in the minimal, six-parameter, standard case is $\Omega_bh^2=0.0226\pm 0.0006$, whereas, after the inclusion of the annihilating DM, becomes $\Omega_bh^2=0.0224 \pm 0.0006$ in the case of \emph{WMAP7} and $\Omega_bh^2=0.0217 \pm 0.0004$ in the case of \emph{CMB ALL+SPT}.  
This lower baryon density required results from the increased number of electrons produced DM annihilations; the two factors combine to give the same optical final depth needed to match the CMB data. 
The constraints on the DM density are only barely affected by the introduction of the DM annihilation, while instead the constraints on the scalar spectral index of primordial perturbations are shifted to higher values.  Similarly to the case of $\Omega_bh^2$, but in the opposite direction, the extra energy injected by the DM annihilation leads to a damping of the tail of CMB power spectrum, so that $n_s$ has to be increased in order to compensate for this effect and still provide a good fit to the data. Note that in the case of \emph{WMAP7}, the introduction of DM annihilation makes the Harrison-Zel'dovich value for the scalar spectral index $n_s=1$ compatible with the data within two standard deviations, while instead when also the SPT data set is added the scale invariant power spectrum is again ruled out by the data. 

\subsection{Simplified energy deposition model}\label{comparison}

As we have stressed already, using a correct description of the energy deposition fractions is crucial to derive reliable DM constraints. Here we intend to quantify this statement by comparing our results with the constraints obtained using an approximated energy deposition model. 

This is summarized by the following expressions: 
\begin{align}
f_{\rm ion,H}& = \tilde{C}_{\rm H} \frac{1-x_{e}}{3} = \tilde{C}_{\rm H} \frac{1 + 2 f_{\rm He} - x}{3(1+2 f_{\rm He})} \label{Eq:Galli1}\\
f_{\rm ion,He}& = \tilde{C}_{\rm He} \frac{1-x_{e}}{3} = \tilde{C}_{\rm He} \frac{1 + 2 f_{\rm He} - x}{3(1+2 f_{\rm He})} \label{Eq:Galli2}\\ 
f_{h}& = \frac{1+2x_{e}}{3} = \frac{(1+2 f_{\rm He})+2x}{3(1+2 f_{\rm He})} \label{Eq:Galli3}
\end{align}
where $x \equiv x_{H} + f_{\rm He} x_{\rm He}$ is a convenient variable to be used in \recfast~and
\begin{equation}
\tilde{C}_{\rm H} = C_{\rm H} + (1-C_{\rm H})\frac{E_{\rm ion,H}}{E_{\alpha,\rm H}}
\end{equation}
(a similar expression is valid for the He) where $C_{\rm H}$ and $C_{\rm He}$ are the Peebles factors as given in~\cite{Wong:2008}.
As in~\cite{Galli:2011} we have multiplied these formulae for the $f_{\rm abs}(z)$ given by~\cite{Slatyer:2009} for the DM annihilation in electrons or muons at 1 GeV.

In Fig.~\ref{Fig:deposition}, we show the corresponding energy depositions as a function of redshift for the muon channel and we compare with what is obtained from the Monte Carlo simulations. It is evident that this simplified approach over predicts the energy deposition for almost the entire redshift range.

We have verified that using the analytic expression in eq.s~\ref{Eq:Galli1}--\ref{Eq:Galli3}, the derived constraints at 1 GeV are found to be coincident with the results reported in table~II by~\cite{Galli:2011} either for the muon or the electron channel. 

\begin{figure}
\centering 
\includegraphics[width=0.49\textwidth]{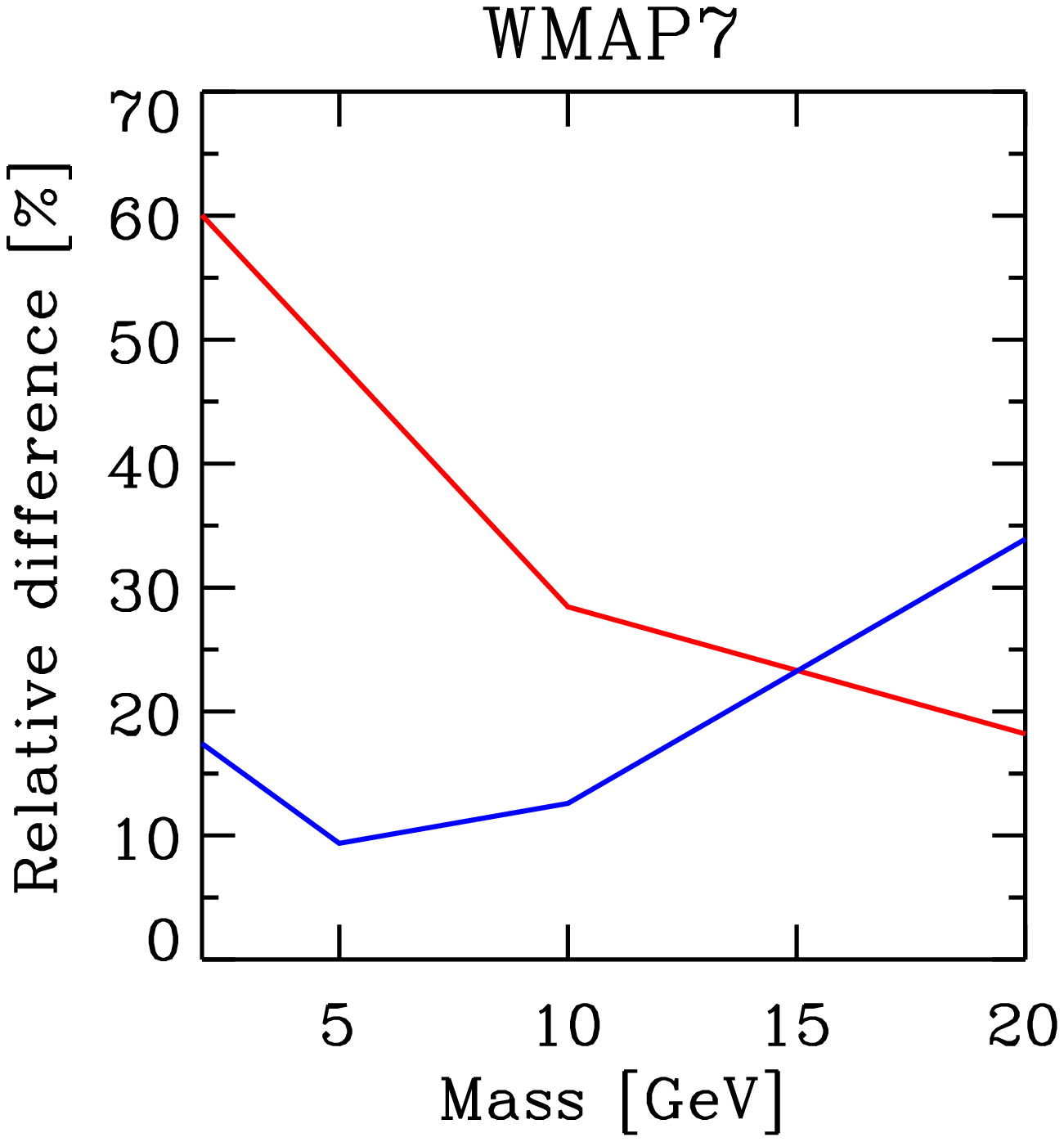}
\caption{Relative differences [(MEDEA-simplified)/MEDEA] between the constraints obtained with a simple energy deposition model as described in Section~\ref{comparison} and using the energy deposition fractions obtained with the \medea~code. Colour code as in Fig.~\ref{Fig:spectra}\label{Fig:comparison}}
\end{figure}

In Fig.~\ref{Fig:comparison}, we show the relative differences between our results and the results obtained adopting the simplified model. We compare the case in which only WMAP7 data are used.
In the range $m_{\rm DM} = 1-20$~GeV, the differences can be quoted between 10 and 30 per cent for the electron channel, and between 20 and 60 per cent for the muon channel. 
The constraints we get always tend to be weaker than those given by~\cite{Galli:2011}: the difference originates from the inclusion of the low-energy processes inducing a net energy-loss (i.e. energy not going into heating, ionization or excitation). As explained in the introduction, decreasing the energy deposition fractions makes the constraints weaker. 

\section{Summary and discussion}

We have investigated the imprints left on the CMB temperature and polarization spectra by the energy deposition due to annihilations of one of the most promising DM candidates, a stable WIMP of mass $m_\chi =1-20$ GeV annihilating into leptons. 
A major improvement with respect to previous similar studies is a detailed treatment of the annihilation cascade and its energy deposition in the cosmic gas.
This is vital as this quantity is degenerate with $\langle \sigma v \rangle$.  

We performed an MCMC analysis using a modified version of the \cosmomc~code and CMB data from the WMAP, QUaD and SPT experiments.
By further marginalizing over the cosmological parameters of the background cosmology, we obtain the constraints on the annihilation cross-section for each annihilation channel.

The strongest constraints are obtained by combining all the available data sets up to $\ell_{\rm max} = 3100$. If annihilation occurs via the $e^+-e^-$ channel, a light WIMP can be excluded as a viable DM candidate in the above mass range. However, if annihilation occurs via $\mu^+-\mu^-$ or $\tau^+-\tau^-$ channel instead, we find that WIMPs with $m_\chi > 5$ GeV cannot be ruled out at 2 $\sigma$ c.l.~to provide the cosmologically required DM content.

We have compared our results with the constraints obtained by assuming a simplified energy deposition model, such as the one profusely used in the recent literature, and we found that realistic energy deposition descriptions can influence the resulting constraints up to 60 per cent. However, at the present stage it was not possible to disentangle the effects of the on-the-spot approximation used in the current analysis from the effects of adopting more realistic low-energy deposition fractions and we postpone to a forthcoming analysis a more detailed comparison between the two approaches.
%A different analysis of the energy deposition uncertainties has been performed in~\citet{Galli:2012}.

We expect that a better understanding of the energy deposition by DM annihilation will be relevant in particular with the upcoming Planck\footnote{The recent analysis of Planck temperature data however gave a weaker constraint than that found from the full WMAP9 temperature + polarization data set because the Planck likelihood does not yet include polarization information at intermediate and high multipoles~\citep{PLANCK2013}.} data, with their better sensitivity, which allow a better constraining of this additional source of ionization.

\section*{Acknowledgements}

We thank C.~L.~Bianco for his precious help.  
We also thank Silvia Galli, Fabio Iocco, Luca Maccione, Pasquale Dario Serpico and Tracy Slatyer for useful comments and discussions.
CE and SP acknowledge a Visiting Grant from SNS where part of this work has been carried out. CE acknowledges support from the Helmholtz Alliance for Astroparticle Physics funded by the Initiative and Networking Fund of the Helmholtz Association. 
The Dark Cosmology Centre is funded by the Danish National Research Foundation. 

\bibliographystyle{mn2e}
\bibliography{medea}

\appendix

\section{Local deposition}\label{Appendix:A}
We have assumed in this paper that photon energy deposition occurs locally, which is not true in general~\citep[see][]{Slatyer:2009}. In the following, we show that this approximation is accurate in the energy range 
of $O(1)$~GeV electrons. CMB photons gain energy as they are inverse Compton scattered by energetic leptons. At each scattering event, a CMB photon with mean energy $E_{\gamma,\rm CMB}$ will be upscattered to an energy equal to:
\begin{equation}\label{Eq:upsca}
E_{\gamma} \approx \frac{4}{3} \gamma^{2} E_{\gamma,\rm CMB} = 0.73 \left( \frac{E_{e}}{\rm GeV} \right)^{2} \left(\frac{1+z}{600}\right) \, \, {\rm MeV} \, ,
\end{equation} 
where $\gamma$ is the Lorentz factor for the lepton. At epochs in which energy deposition is important ($z \le 1000$) such upscattered photons are subsequently mainly downgraded by Compton scattering with thermal electrons~\citep{Chen:2004,Slatyer:2009}. 

\begin{figure}
\centering 
\includegraphics[width=0.49\textwidth]{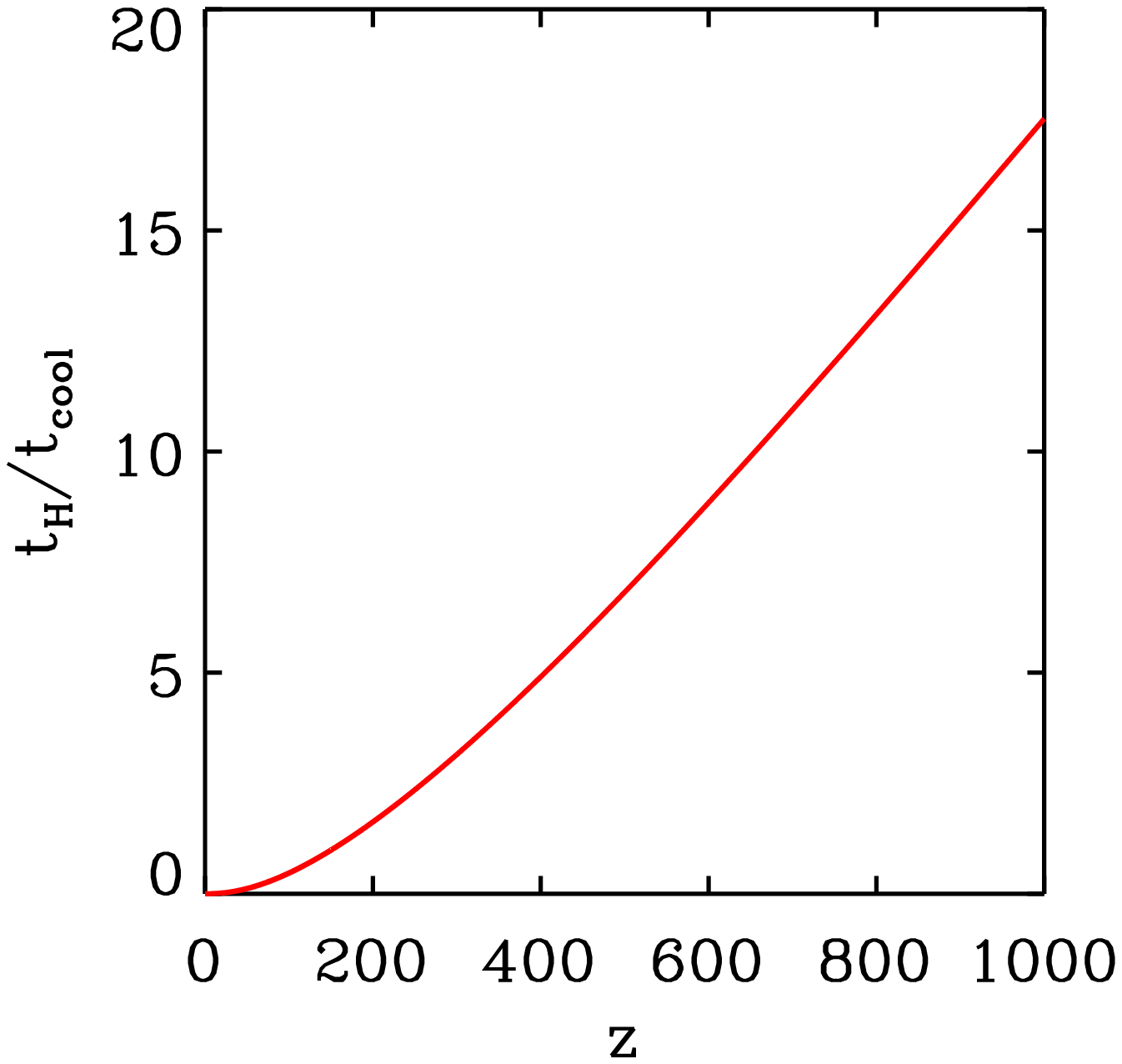}
\caption{Ratio between Hubble time and Compton cooling time-scales for the photon upscattered by a 1 GeV electron as a function of the redshift. \label{Fig:compton}}
\end{figure}

To estimate the efficiency of this mechanism, we compare in Fig.~\ref{Fig:compton} the Compton cooling time, $t_{\rm cool}^{-1} = (dlnE/dt) = c\sigma_{T} n_{e} (1+z)^{3} \epsilon g(\epsilon)$, of a photon upscattered by a 1 GeV electron (see equation~\ref{Eq:upsca}) with the Hubble time, $t_{H} = H^{-1} (z)$, where $\epsilon \equiv E_{\gamma}/m_{e}c^{2}$ and $g(\epsilon)$ is the classical Klein-Nishina cross-section. 
It is evident that the local deposition assumption, requiring  $t_{H} \gg t_{\rm cool}$, can be safely applied in the redshift range of interest here ($z \sim 600$ as confirmed by the principal component analysis performed by~\citet{Finkbeiner:2012}). Note however that ICS produces a broad spectrum of photons, and photons produced at lower energies with respect to the peak energy of equation~\ref{Eq:upsca} cool more slowly than their higher energy counterparts. This can in part explain the differences between the results we obtain with the two deposition models described in Section~\ref{comparison}.

\section{Fitting formulae}

Below are the fitting formulae to the numerically derived energy depositions of electrons and positrons 
in the various channels.
\begin{eqnarray}
f_{h}(x_{e},z) & = & 10^{A(z)}(1-C(z) (1-x_{e}^{B(z)})) \label{Eq:fit1}\\
f_{ion,H}(x_{e},z) & = &  10^{A(z)}(1-x^{B(z)})^{C(z)} \label{Eq:fit2}\\
f_{ion,He}(x_{e},z) & = &  10^{A(z)}(1-x^{B(z)})^{C(z)} \label{Eq:fit3}
\end{eqnarray}
where:
\begin{eqnarray}
A(z) & = & A_{0} + A_{1}  \log_{10} z + A_{2}  (\log_{10} z)^{2} \\ 
B(z) & = & B_{0} + B_{1}  \log_{10} z + B_{2}  (\log_{10} z)^{2} \\ 
C(z) & = & C_{0} + C_{1}  \log_{10} z + C_{2}  (\log_{10} z)^{2} 
\end{eqnarray}
The values of the parameters are given in Tab.~\ref{Tab:ee}.

Moreover, we provide an updated version with respect to~\citet{Evoli:2012} for the energy deposition fractions that can be used for energies below the IC threshold: $E_{\rm th} = ((1+z)/21)^{-1/2}$~MeV:
\begin{eqnarray}
f_{h}(x_{e}) & = & a (1-c(1-x_{e}^{b})) \label{Eq:low1}\\
f_{\rm ion}(x_{e}) & = & a (1-x_{e}^{b})^{c} \label{Eq:low2}
\end{eqnarray}
where the values of the parameters are given in Tab.~\ref{Tab:low}.

\begin{table*}
\caption{Parameter values to be used in equations~\ref{Eq:fit1} and \ref{Eq:fit3}}
\begin{center}
\small
\begin{tabular}{|c|c|c|c|c|c|c|c|c|c|c|c|}
\hline
$f_{i}$ & Channel & Mass & $A_{0}$ & $A_{1}$ & $A_{2}$ & $B_{0}$ & $B_{1}$ & $B_{2}$ & $C_{0}$ & $C_{1}$ & $C_{2}$ \\
\hline
$f_{\rm ion,H}$ & $\mu^+\mu^-$ & 2 &  -1.996 & 1.668e-3 & -1.049e-6 & 3.750e-1 & -2.778e-5 & -3.301e-8 & 9.643e-1 & 2.912e-5 & -3.610e-7\\
$f_{\rm ion,H}$ & $e^+e^-$         & 2 &  -1.405 & 2.357e-3 & -1.549e-6 & 3.820e-1 & -1.612e-4 & 8.698e-8 & 1.014 & -7.705e-4 & 3.240e-7\\
$f_{\rm ion,H}$ & $\mu^+\mu^-$ & 5 &  -1.937 &  2.538e-3 & -1.664e-6 & 3.790e-1 & -1.901e-4 & 1.225e-7 & 9.904e-1 & -9.910e-4 & 6.400e-7\\
$f_{\rm ion,H}$ & $e^+e^-$         & 5 &  -1.678 &  3.279e-3 & -2.211e-6 & 3.739e-1 & -2.203e-4 & 1.744e-7 & 9.649e-1 & -1.284e-3 & 1.038e-6\\
$f_{\rm ion,H}$ & $\tau^+\tau^-$ & 5 &  -1.853 &  1.758e-3 & -1.093e-6 & 3.732e-1 & -7.861e-5 & 2.318e-8 & 9.549e-1 & -3.364e-4 & 4.671e-8\\
$f_{\rm ion,H}$ & $\mu^+\mu^-$ & 10 &  -2.053 &  2.927e-3 & -1.944e-6 & 3.704e-1 & -1.873e-4 & 1.408e-7 & 9.551e-1 & -1.064e-3 & 8.178e-7 \\
$f_{\rm ion,H}$ & $e^+e^-$         & 10 &  -1.888 &  3.890e-3 & -2.648e-6 & 3.673e-1 & -1.465e-4 & 1.159e-7 & 8.935e-1 & -8.101e-4 & 7.575e-7 \\
$f_{\rm ion,H}$ & $\tau^+\tau^-$ & 10 &  -2.117 &  1.975e-3 & -1.245e-6 & 3.709e-1 & -9.739e-5 & 4.789e-8 & 9.459e-1 & -4.860e-4 & 2.286e-7 \\
$f_{\rm ion,H}$ & $\mu^+\mu^-$ & 20 &  -2.256 &  3.536e-3 & -2.383e-6 & 3.618e-1 & -1.410e-4 & 1.246e-7 & 8.983e-1 & -8.122e-4 & 7.281e-7\\
$f_{\rm ion,H}$ & $e^+e^-$         & 20 &  -2.100 &  4.393e-3 & -2.980e-6 & 3.784e-1 & -4.212e-5 & -1.377e-8 & 8.560e-1 & -3.335e-5 & -3.348e-9\\
$f_{\rm ion,H}$ & $\tau^+\tau^-$ & 20 &  -2.317 &  2.608e-3 & -1.699e-6 & 3.684e-1 & -1.296e-4 & 9.050e-8 & 9.334e-1 & -7.095e-4 & 5.067e-7\\
$f_{\rm ion,He}$ & $\mu^+\mu^-$ & 2 & -2.521 & 9.046e-4 &-3.260e-7  & 4.445e-1 & 1.200e-4  & -1.484e-7 & 4.350e-1 &  1.148e-3 & -1.260e-6\\
$f_{\rm ion,He}$ & $e^+e^-$         & 2 & -1.978   & 1.916e-3 & -1.026e-6 & 4.631e-1 & -4.478e-5 & 1.766e-8  & 5.335e-1 &  2.848e-4 &-5.717e-7\\
$f_{\rm ion,He}$ & $\mu^+\mu^-$ & 5 & -2.514   & 2.391e-3 & -1.445e-6 & 4.579e-1 & -1.085e-4 & 6.551e-8  & 5.133e-1 & -1.957e-4 & -3.395e-8\\
$f_{\rm ion,He}$ & $e^+e^-$         & 5 & -2.286    & 3.561e-3 & -2.399e-6 & 4.558e-1 & -1.212e-4 & 9.403e-8  & 5.343e-1 & -6.910e-4 & 4.795e-7\\
$f_{\rm ion,He}$ & $\tau^+\tau^-$ & 5 & -2.386    & 1.252e-3 & -6.014e-7 & 4.441e-1 & 2.714e-5  & -5.451e-8 & 4.389e-1 & 5.135e-4  & -6.285e-7\\
$f_{\rm ion,He}$ & $\mu^+\mu^-$ & 10 & -2.637  & 2.994e-3 & -1.949e-6 & 4.444e-1 & -1.203e-4 & 9.085e-8  & 4.894e-1 & -4.346e-4 & 2.701e-7\\
$f_{\rm ion,He}$ & $e^+e^-$         & 10 & -2.468 & 4.217e-3 & -2.977e-6 & 4.687e-1 & 2.201e-5  & -1.061e-7 & 5.021e-1 & -4.570e-4 & 3.646e-7\\
$f_{\rm ion,He}$ & $\tau^+\tau^-$ & 10 & -2.656 & 1.611e-3 & -8.855e-7 & 4.410e-1 & -5.718e-6 & -1.931e-8 & 4.377e-1 & 2.615e-4  & -3.665e-7\\
$f_{\rm ion,He}$ & $\mu^+\mu^-$ & 20 & -2.830  & 3.732e-3 & -2.569e-6 & 4.432e-1 & -8.121e-5 & 7.030e-8  & 4.619e-1 & -4.183e-4 & 3.644e-7\\
$f_{\rm ion,He}$ & $e^+e^-$         & 20 & -2.634  & 4.431e-3 & -3.097e-6 & 5.187e-1 & -2.716e-4 & 1.468e-7  & 4.977e-1 & -2.768e-4 & 2.312e-7\\
$f_{\rm ion,He}$ & $\tau^+\tau^-$ & 20 & -2.872  & 2.494e-3 & -1.573e-6 & 4.418e-1 & -4.527e-5 & 2.369e-8  & 4.490e-1 & -8.283e-5 & -1.347e-8\\
$f_{h}$ &$\mu^+\mu^-$ & 2  & -1.621  & 1.851e-3 & -1.269e-6 & 2.625e-1 &  1.347e-4 & -1.029e-7 & 8.849e-1 & -3.992e-5 &  1.456e-8\\
$f_{h}$ &$e^+e^-$         & 2  & -1.013 & 2.349e-3 & -1.624e-6 & 2.670e-1 &  1.418e-4 & -1.169e-7 & 8.860e-1 & -8.014e-5 &  4.808e-8\\
$f_{h}$ &$\mu^+\mu^-$ & 5  & -1.546  & 2.419e-3 & -1.606e-6 & 2.690e-1  &  1.011e-4 & -8.034e-8 & 8.847e-1 & -8.490e-5 &  6.019e-8\\
$f_{h}$ &$e^+e^-$         & 5  & -1.279 &  2.948e-3 & -1.926e-6 & 2.794e-1 &  2.208e-5 & -9.847e-9 & 8.807e-1 & -9.096e-5 &  7.457e-8\\
$f_{h}$ &$\tau^+\tau^-$ & 5  & -1.476  &  1.828e-3 & -1.197e-6 & 2.651e-1 &  1.146e-4 & -8.667e-8 & 8.839e-1 & -5.171e-5 &  2.939e-8\\
$f_{h}$ &$\mu^+\mu^-$ & 10 & -1.665& 2.735e-3 & -1.796e-6 & 2.734e-1 & 6.668e-5 & -5.288e-8 & 8.815e-1 & -8.244e-5 &  6.480e-8\\
$f_{h}$ &$e^+e^-$         & 10 & -1.516&  3.609e-3 & -2.362e-6 & 2.800e-1 & -7.608e-5 &  8.341e-8 & 8.719e-1 & -5.938e-6 &  3.950e-9\\
$f_{h}$ &$\tau^+\tau^-$ & 10 & -1.739&  1.989e-3 & -1.290e-6 & 2.668e-1 & 1.019e-4 & -7.745e-8 & 8.830e-1& -5.687e-5 & 3.657e-8\\
$f_{h}$ &$\mu^+\mu^-$ & 20 & -1.881& 3.307e-3 & -2.166e-6 & 2.765e-1 & -1.291e-5 &  1.540e-8 & 8.763e-1 & -4.793e-5 &  4.419e-8\\
$f_{h}$ &$e^+e^-$         & 20 & -1.767&  4.314e-3 & -2.857e-6 & 2.580e-1 & -6.141e-5 &  1.013e-7 & 8.787e-1 & 1.085e-5 & -2.599e-8\\
$f_{h}$ &$\tau^+\tau^-$ & 20 & -1.938 &   2.507e-3 & -1.627e-6  & 2.703e-1 &  6.107e-5 & -4.581e-8 & 8.812e-1 & -6.110e-5 &  4.604e-8\\

\hline
\end{tabular}
\end{center}
\label{Tab:ee}
\end{table*}

\begin{table*}
\caption{Parameter values to be used in equations~\ref{Eq:low1} and \ref{Eq:low2}}
\begin{center}
\small
\begin{tabular}{|c|c|c|c|}
\hline
$f_{i}$ & $a$ & $b$ & $c$ \\
\hline
$f_{h}$ & 9.77e-1 & 3.00e-1 & 9.00e-1\\
%f_lyman & = & fi(xe, 3.51E-01,  2.90E-01,  1.08E+00)\\
$f_{\rm ion,H}$ & 3.55e-1 & 3.90e-1 & 1.11\\
$f_{\rm ion,He}$ & 6.10e-2 & 5.30e-1 & 1.05\\
\hline
\end{tabular}
\end{center}
\label{Tab:low}
\end{table*}

\label{lastpage}
\end{document}